\documentclass[aps,prl,twocolumn,superscriptaddress]{revtex4}

\usepackage{graphicx}
\usepackage{dcolumn}
\usepackage{bm}
\usepackage{color}

\begin{document}
\title{Observation of a two-dimensional spin-lattice in
non-magnetic semiconductor heterostructures}
\author{Christoph Siegert}
\affiliation{Cavendish Laboratory, University of Cambridge, J.J.
Thomson Avenue, Cambridge CB3 0HE, United Kingdom.}
\author{Arindam Ghosh}
\email{arindam@physics.iisc.ernet.in} \affiliation{Cavendish
Laboratory, University of Cambridge, J.J. Thomson Avenue, Cambridge
CB3 0HE, United Kingdom.} \affiliation{Department of Physics, Indian
Institute of Science, Bangalore 560 012, India.}
\author{Michael Pepper}
\affiliation{Cavendish Laboratory, University of Cambridge, J.J.
Thomson Avenue, Cambridge CB3 0HE, United Kingdom.}
\author{Ian Farrer}
\affiliation{Cavendish Laboratory, University of Cambridge, J.J.
Thomson Avenue, Cambridge CB3 0HE, United Kingdom.}
\author{David A. Ritchie}
\affiliation{Cavendish Laboratory, University of Cambridge, J.J.
Thomson Avenue, Cambridge CB3 0HE, United Kingdom.}
\date{\today}
\maketitle

{\bf Tunable magnetic interactions in high-mobility nonmagnetic
semiconductor heterostructures are centrally important to spin-based
quantum technologies. Conventionally, this requires incorporation of
``magnetic impurities'' within the two-dimensional (2D) electron
layer of the heterostructures, which is achieved either by doping
with ferromagnetic atoms~\cite{ohno1}, or by electrostatically
printing artificial atoms or quantum
dots~\cite{goldhaber-gordon,cronenwett,jeong,craig_marcus}. Here we
report experimental evidence of a third, and intrinsic, source of
localized spins in high-mobility GaAs/AlGaAs heterostructures, which
are clearly observed in the limit of large setback distance (${\rm
\approx 80}$ nm) in modulation doping. Local nonequilibrium
transport spectroscopy in these systems reveal existence of multiple
spins, which are located in a quasiregular manner in the 2D Fermi
sea, and mutually interact at temperatures below 100 milliKelvin via
the Ruderman-Kittel-Kasuya-Yosida (RKKY) indirect exchange. The
presence of such a spin-array, whose microscopic origin appears to
be disorder-bound, simulates a 2D lattice-Kondo system with
gate-tunable energy scales.}

Unintentional magnetic impurities are expected to be absent in
high-quality nonmagnetic semiconductors, such as molecular beam
epitaxy-grown GaAs/AlGaAs heterostructures. Contrary to this general
belief, recent observation of a Kondo-like resonance in low-energy
density-of-states of one dimensional (1D) quantum wires indicated
existence of localized spin in mesoscopic GaAs/AlGaAs-based
devices~\cite{cronenwett_1D}. Subsequently, evidence of localized
spins was also reported in unconfined quasi-ballistic 2D
systems~\cite{selfPRL_04,selfPRL_05}, instigating the question
whether localized spins are intrinsic to GaAs/AlGaAs-based
nonmagnetic semiconductors, and if so, what is the microscopic
origin of these spins. A fascinating aspect of this problem is the
possibility of a layer of mutually interacting spins: a system of
considerable importance in studying various forms of magnetic
ordering, quantum phase transitions and non-Fermi liquid effects
(see Ref.~\cite{stewart} for a review).

In the past experiments with ballistic 1D or 2D
systems~\cite{cronenwett_1D,selfPRL_04,selfPRL_05},
localized spins were detected with nonequilibrium transport
spectroscopy. Pronounced structures, commonly known as zero-bias
anomaly (ZBA), in differential conductance ($dI/dV$) of mesoscopic
devices close to zero source-to-drain electric potential ($V_{\rm
SD}$) were interpreted to be a consequence of spin-spin or
spin-conduction electron exchange interaction. Here, we have
augmented the nonequilibrium transport spectroscopy with perpendicular-field
magnetoresistance measurements, which not only confirm the existence
of multiple localized spins within high-mobility GaAs/AlGaAs
heterostructures, but also reveals a striking order in the spatial
distribution of these spins that becomes visible over a narrow range
of $n_{\rm 2D}$.

Mesoscopic devices fabricated from Si monolayer-doped GaAs/AlGaAs
heterostructures were used, where the 2D electron layer was formed
300 nm below the surface. A thick ($\approx 80$ nm) spacer layer of
undoped AlGaAs between the dopants and the electrons provided a
heavily compensated dopant layer with a filling factor $f \approx
0.9$. The resulting high electron mobility ($\sim 1-3\times10^6$
cm$^2$/V-s) provides a long as-grown elastic mean free path $\sim 6
- 8$ $\mu$m, which acts as an upper limit to the device dimensions,
ensuring quasi-ballistic transport. (Micrograph of a typical device
is shown in Fig.~1a.)

At low electron temperatures ($T \lesssim 100$ mK) and zero magnetic
field, both equilibrium and nonequilibrium transport display rich
structures well up to linear conductance $G \sim
10-15\times(e^2/h)$, as the voltage $V_{\rm G}$ on the gate is
increased. For most devices the structures are strongest at carrier
density $n_{2D} \sim 1-3 \times 10^{10} cm^{-2}$ (Fig.~1b), and
consist of a repetitive sequence of two-types of resonances at the
Fermi energy ($E_{\rm F}$). This is illustrated in the surface plot
of $dI/dV$ in Figs.~1c for device~1 of Fig.~1a. We denote the strong
single-peak resonance at $V_{\rm SD} = 0$ as ZBA-I, which splits
intermittently to form a double-peaked ZBA with a gap at $E_F$,
henceforth referred to as ZBA-II. The illustrations of ZBA-I and
ZBA-II in the inset of Fig.~1c were recorded at points I and II in
Fig~1b, respectively. We define $\Delta$ as the half-width at
half-depth of ZBA-II. While similar nonequilibrium characteristics
was observed in over 50 mesoscopic devices from 5 different wafers,
reducing the setback distance below $\sim 60 - 80$ nm was generally
found to have detrimental effect on the clarity of the resonance
structures (for both ZBA-I and ZBA-II), often leading to broadening
or complete suppression.

The nature of suppression of such low-energy resonances with
increasing $T$ and at finite in-plane magnetic field ($B_{||}$)
indicate Kondo-like exchange (Figs.~1d-1g)~\cite{selfPRL_05}, and
hence presence of localized moments. A complete description can be
obtained with the so-called ``two impurity'' Kondo model, which
embodies the interaction of an ensemble of localized spins within
the sea of conduction electrons
~\cite{jayaprakash_HRK,affleck,pustilnik,vavilov_glaz,golovach_loss,hofstetter}.
(See Ref.~\cite{selfPRL_05} and Supplementary Information for
arguments against alternative explanation of the ZBA.) In the
presence of antiferromagnetic coupling of individual spins to
surrounding conduction electrons, the zero-field splitting of the
Kondo-resonance (ZBA-II) arises due to a $V_{\rm G}$-dependent,
oscillatory inter-impurity exchange $J_{12}$, leading to the gap
$\Delta \sim |J_{12}|$ at $E_{\rm F}$. At certain intermediate
values of $V_{\rm G}$ one obtains ZBA-I when $|J_{12}| \ll k_{\rm
B}T$. For ZBA-II, nonzero $J_{12}$ results in a nonmonotonic
suppression of $dI/dV$ at low bias ($|V_{\rm SD}|  \lesssim \Delta$,
as indeed observed experimentally (Figs.~1e and 1g), while for ZBA-I
this decrease is monotonic, and reflects suppression of
single-impurity Kondo-resonance at individual noninteracting spins
(see Figs.~1d and 1f)~\cite{selfPRL_05}.

Mesoscopic devices showing clear resonances in the nonequilibrium
transport also display a characteristic linear magnetoresistance
(MR) over the same range of $n_{\rm 2D}$, when a small magnetic
field ($B_\perp$) is applied $perpendicular$ to the plane of 2D
electron layer. As shown for a different device, at high T ($\cong
1.4$~K), the MR consists of small peak-like structures at specific
values of $B_\perp$ superposed on a parabolically increasing
background (Fig.~2d), while at low T ($\cong 30$~mK), the
magnetotransport breaks into quasi-periodic oscillations
irrespective of the structure of the ZBA (Fig.~2a and 2b). We note
that both these evidences indicate electron transport in a
mesoscopic, quasi-regular array of $antidots$, which has been
studied extensively in artificially fabricated
antidot-lattices~\cite{weiss1,schuster,weiss}. In the classical
regime (high $T$), the peaks in MR correspond to commensurable
cyclotron orbits enclosing fixed number of antidots ~\cite{weiss1},
while at low $T$, quantum interference leads to the phase coherent
oscillations in magnetoconductance, arising from transport along
multiple connected Aharonov-Bohm rings as the inelastic scattering
length exceeds sample dimensions ~\cite{schuster}.

The magnetotransport data shown in Fig.~2 allows an estimation of
the inter-antidot distance $R$. In Fig.~2d, on subtracting the
background, signature of commensurable orbits at $B_\perp \approx$
0.08, 0.05 and 0.03 T corresponding to cyclotron radius of the
electron encircling one, two and four antidots respectively (see
inset). While this gives $R \sim 500$~nm (using the $n_{\rm 2D}
\approx 1.31 \times 10^{10}$ cm$^{-2}$), a more accurate estimate of
$R$ was obtained by Fourier transforming the low-T phase-coherent
oscillations of Fig.~2a and 2b. In Fig.~2c, the power spectra
calculated over the range 0 to 0.065 T at both values of $V_{\rm G}$
show a strong peak at the frequency $f_{\rm \Delta B} \approx eR^2/h
\approx 105 \pm 10$ T$^{-1}$, corresponding to one flux quantum
through unit cell of the antidot lattice (orbit b). This gives $R
\approx 670 \pm 30$ nm, which is consistent with the estimate from
commensurability effect. Satellite peaks often appears in the power
spectra, for example those at $f_{\rm \Delta B} \cong 190$ T$^{-1}$
(orbit c) and 50 T$^{-1}$ (orbit a), which can be associated to
specific stable orbits as indicated in the inset of Fig.~2a. $R$ was
found to be weakly device-dependent, varying between $600 - 800$ nm,
but insensitive to the lithographic dimensions of the devices.

Collectively, the observed nonequilibrium characteristics and
low-field MR results indicate formation of a quasi-regular 2D
spin-lattice embedded within the Fermi sea, where apart from the
Kondo-coupling, the conduction electrons would also undergo
potential scattering at the lattice sites. The nature of such
potential scattering can be cotunneling~\cite{pustilnik}, or
scattering off the tunnel barrier at the localized
sites~\cite{wolf}. In this framework, the RKKY exchange between the
spins naturally leads to an oscillatory behavior of $J_{12}$, where
range function $\Psi(2k_{\rm F}R)$ in the interaction magnitude
reverses its sign with a periodicity of $\pi$ in $2k_{\rm F}R$,
$k_{\rm F} = \sqrt{2 \pi n_{2D}}$ being the Fermi wave vector.
Analytically~\cite{beal-monod},

\begin{equation}
\label{rkky} \Delta \sim |J| \sim E_{\rm F}(J\rho_{\rm
2D})^2|\Psi(2k_{\rm F}R)|
\end{equation}

\noindent where $\rho_{\rm 2D}$ is the 2D density of states and $J$
is the exchange coupling between an impurity spin and local
conduction electron. Fig.~3a shows the direct confirmation of this,
where we have plotted $\Delta$ as a function of $2k_{\rm F}R$ for
the device in Fig.~1c. The clear periodicity of $\approx \pi$
(within $\pm 5\%$) in $2k_{\rm F}R$ can be immediately recognized as
the so-called ``$2k_{\rm F}R$-oscillations'' in the RKKY
interaction, establishing the spin-lattice picture.

The absolute magnitude of $J_{12}$, and hence $\Delta$, for a 2D
distribution of spins may differ widely from the simple two-impurity
RKKY interaction, and would be affected by frustrated magnetic
ordering or spin glass freezing~\cite{Hirsch}, as well as deviation
from perfect periodicity in the spin arrangements~\cite{roche}.
Nevertheless, a framework for relative comparison of $\Delta$ in
different samples can be obtained from Eq.~\ref{rkky} by normalizing
$\Delta$ with $E_{\rm F}$. As shown in Fig.~3b, adjusting for the
experimental uncertainty in $k_{\rm F}$ and $R$, $\Delta/E_{\rm F}$
for four different devices with various lithographic dimensions can
be made to collapse on the solid line proportional to modulus of
pairwise RKKY range function over a wide range of $2k_{\rm
F}R$~\cite{beal-monod}.

With known inter-spin distance, we shall now discuss two outstanding
issues of this paper: (1) the microscopic origin of the uniform
array of antidots, and subsequently, (2) emergence of the localized
spins. A structurally intrinsic origin of antidots in 2D Fermi sea
of modulation-doped high-mobility GaAs/AlGaAs heterostructures can
arise from long-range potential fluctuations in the conduction band.
Direct experimental imaging of disorder in similar systems reported
typical distance between fluctuation to be $\sim 0.5 - 1$~$\mu$m, in
excellent agreement to the magnitude of $R$ in our
devices~\cite{finkelstein}. In presence of strong correlation in the
dopant layer at large $f$, theoretical investigations have also
indicated a well-defined length scale in the spatial distribution of
the potential fluctuations~\cite{dohler}.

A disorder-templated localized moment formation can then be
envisaged through local depletion of electrons, analogous to  moment
formation in metal-semiconductor Schottky barriers~\cite{wolf}. We
discuss this on the basis of three common representations of
disorder and screening in high-mobility GaAs/AlGaAs systems, as
schematized in Fig.~4a-c~\cite{efros}. At high $n_{\rm 2D}$, the
background disorder is linearly screened at all points, with local
charge fluctuations $|\delta n_{\rm 2D}| \ll \langle n_{\rm
2D}\rangle$ (Fig.~4a). With decreasing $n_{\rm 2D}$, linear
screening will break down {\it locally} at the maxima of slow
potential fluctuations, where $n_{\rm 2D}$ becomes smaller than the
local (rapid) density fluctuations of the dopants, resulting in the
formation of single-particle localized states. Assuming a random
distribution of the dopants, this is expected to occur at $n_{\rm
2D}^{\rm c} \sim [(1-f)n_\delta/\pi]^{1/2}/\xi$, where $n_\delta =
2.5\times 10^{12}$ cm$^{-2}$ is the bare dopant density in our
devices, and $\xi$ is the localization length. From $f_{\rm \Delta
B}$ of orbit a in Fig.~2a we estimate $\xi \sim 150$ nm, which gives
$n_{\rm 2D}^{\rm c} \approx 1.9\times 10^{10}$ cm$^{-2}$, which
indeed marks the onset of strong ZBA (see Fig.~1b).

On further lowering of $n_{\rm 2D}$, the system crosses over to the
strongly localized regime ($G \ll e^2/h$), and the 2D electron
system disintegrates into puddles, which are often interconnected
through quantum point contacts (Fig.~4c). While local spins can form
at the point contacts~\cite{cronenwett_1D,graham}, the commensurability
effect and phase-coherent magnetoconductance oscillations shown in
Fig.~2 cannot be explained in such a picture, as they require
extended and uninterrupted electron orbits.

Our experiments thus outline a new microscopic mechanism of local
moment formation in high-mobility GaAs/AlGaAs-based semiconductors
with remote modulation doping. This has serious implications on the
possibility of a gate-tunable $static$ spontaneous spin polarization
in mesoscopic devices at low temperatures, and hinged on the
Kondo-coupling of the localized moments to the surrounding
conduction electrons. To verify such a coupling, we have estimated
the ratio $\epsilon/\Gamma$ for localized states using
experimentally observed Kondo temperature $T_{\rm K}$, and that
$T_{\rm K} \sim (E_{\rm F}/k_{\rm B})\exp(\pi\epsilon/2\Gamma)$ in
the $U \rightarrow \infty$ limit, where $\epsilon$ and $\Gamma$ are
the energy of single-electron state (with respect to $E_{\rm F}$),
and level broadening respectively (Fig.~4b), and $U \sim e^2/\xi$
$\gg E_{\rm F}$, is the on-site Coulomb repulsion. Taking the
measured $T_{\rm K} \approx 265$~mK at ZBA-I at point I in Fig.~1c
as an example, we find $\epsilon/\Gamma \approx -2.1$, which
confirms the Kondo regime.

\noindent \textbf{Acknowledgement:} We acknowledge discussions with C. J. B. Ford,
G. Gumbs, M. Stopa, P. B. Littlewood, H. R. Krishnamurthy, B. D.
Simons, C. M. Marcus, D. Goldhaber-Gordon and K. F. Berggren. The
work was supported by an EPSRC funded project. C.S. acknowledges
financial support from Gottlieb Daimler- and Karl Benz-Foundation.

\newpage
\begin{figure*}[t]
\centering
\includegraphics[height=5.97cm,width=15cm]{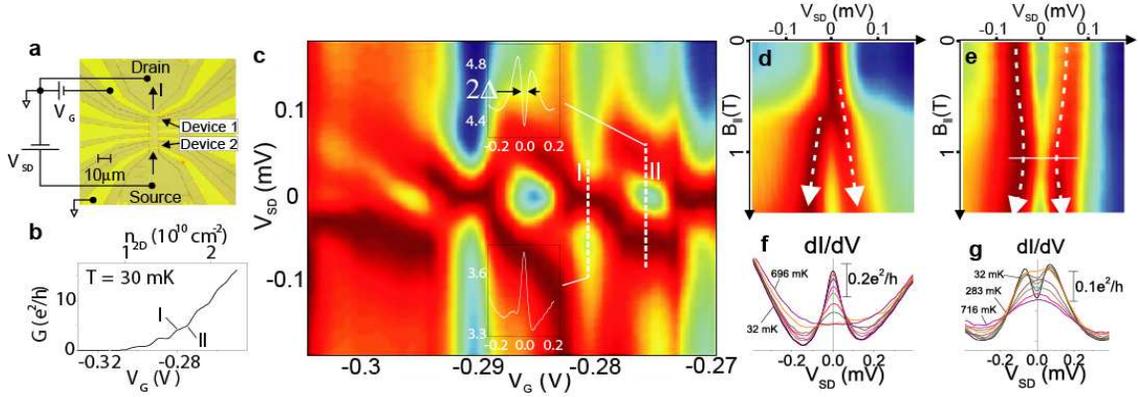}
\caption{Nonequilibrium characteristics: {\bf a,} Picture of a set
of two typical devices and electrical connections. Active area of a
device is defined by the gate-covered region of the etched mesa. The
data shown here was obtained from device 1, with device 2 and other
side-gates kept grounded. {\bf b,} Typical linear conductance, G,
vs. gate voltage $V_{\rm G}$ (and electron density $n_{\rm 2D}$) for
device~1 at 30 mK and zero external magnetic field. {\bf c,} Surface
plot of differential conductance $dI/dV$ of device~1 in $V_{\rm SD}
- V_{\rm G}$ plain. Each $dI/dV-V_{\rm SD}$ trace at a particular
$V_{\rm G}$ was vertically shifted for leveling. The insets
illustrate ZBA-I and ZBA-II-type resonances at points I and II in
Fig.~1b, respectively. {\bf d,} Surface plot of ZBA-I in in-plane
magnetic field ($B_{||}$). The dashed line shows a linear monotonic
splitting with effective $g$-factor $|g^*| \approx 0.5$, which
confirms the role of spin. {\bf e,} Surface plot of ZBA-II in
$B_{||}$. The single-impurity Kondo-behavior dominates above $B_{||}
\sim \Delta/g^\ast\mu_{\rm B}$ (the horizontal line), where $\Delta$
is the half-gap defined in text. {\bf f,} Monotonic temperature
($T$) suppression of ZBA-I. {\bf g,} $T$-dependence of $dI/dV$ at
ZBA-II. $dI/dV$ is nonmonotonic in $T$ for $|V_{\rm SD}| \lesssim
\Delta$.}
\end{figure*}

\newpage
\begin{figure*}[t]
\centering
\includegraphics[height=7.84cm,width=15cm]{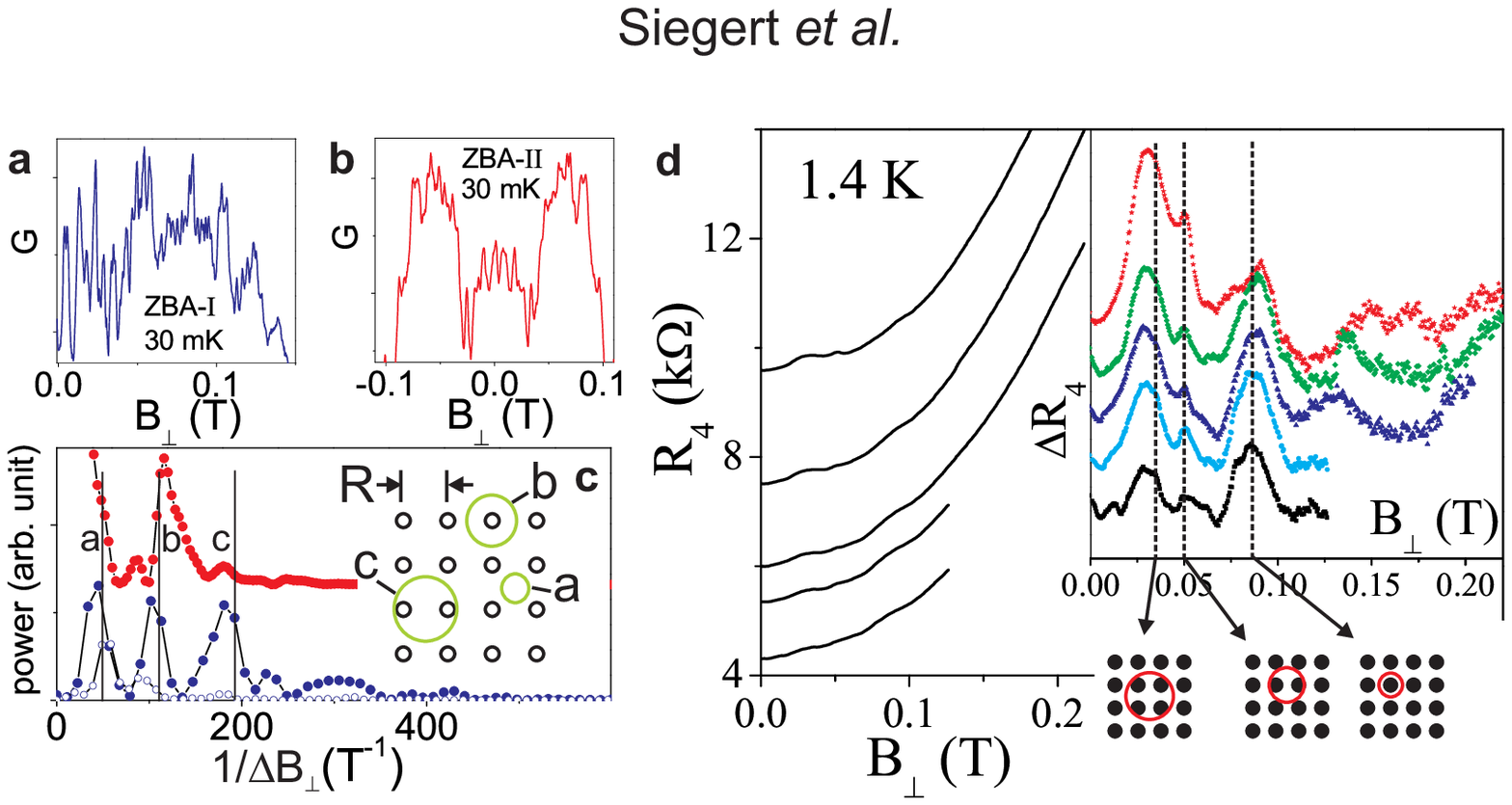}
\caption{Quantum and classical magnetotransport in perpendicular
magnetic field ($B_\perp$). Typical linear magnetoconductance
oscillation at {\bf a,} a single-peak resonance (ZBA-I), and {\bf
b,} a double-peak resonance (ZBA-II). {\bf c,} Power spectra of the
magnetoconductance oscillations. The filled markers (blue: ZBA-I and
red: ZBA-II) represent spectra obtained from the range $|B_\perp|
\leq 0.065$~T, while the empty (blue) marker represents the spectrum
(vertically scaled for clarity) from $0.065$~T $< B_\perp < 0.15$~T.
The orbits corresponding to the peaks are indicated in the
schematic. {\bf d,} Four-probe linear magnetoresistance at 1.4 K for
five electron densities from 1.22 (topmost trace) to $1.39\times
10^{10}$ cm$^{-2}$ (bottom trace), where well-defined ZBA's appear
at low temperatures. Inset: Magnetoresistance after subtracting the
parabolic background. The dashed lines, which denote various
commensurate orbits, are computed using the average density of
$1.31\times10^{10}$ cm$^{-2}$, and $R \approx 500$~nm.}
\end{figure*}

\newpage
\begin{figure*}[t]
\centering
\includegraphics[height=10.14cm,width=10cm]{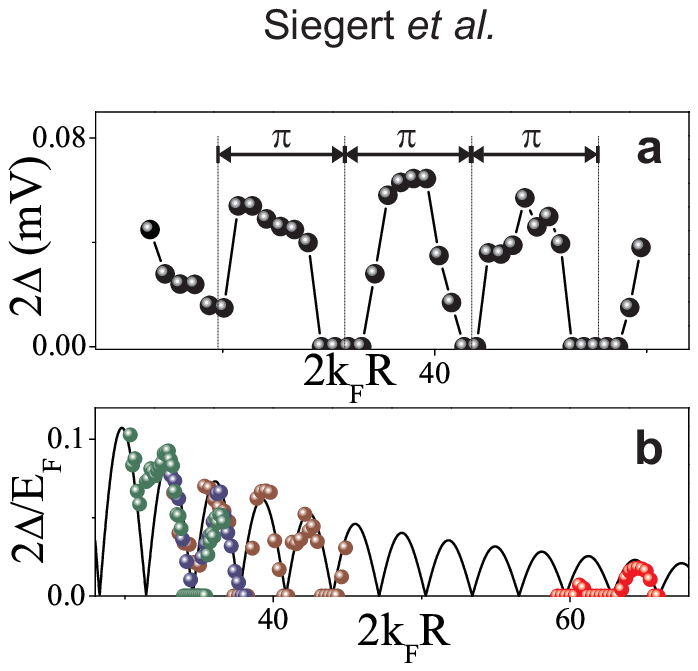}
\caption{RKKY indirect exchange and ``$2k_{\rm F}R$-oscillations'':
{\bf a,} $2\Delta$ from Fig.~1c as a function $2k_{\rm F}R$, where
$k_{\rm F}$ is the Fermi wave vector, and $R$ is the inter-spin
distance obtained from magnetoconductance oscillations. {\bf b,}
$2\Delta/E_{\rm F}$ as a function of $2k_{\rm F}R$ for four
different devices. Solid line is proportional to the pairwise RKKY
range function (see text). Local disorder governs the experimentally
attainable range of $2k_{\rm F}R$ within a given mesoscopic device.}
\end{figure*}

\newpage
\begin{figure*}[t]
\centering
\includegraphics[height=16.68cm,width=7cm]{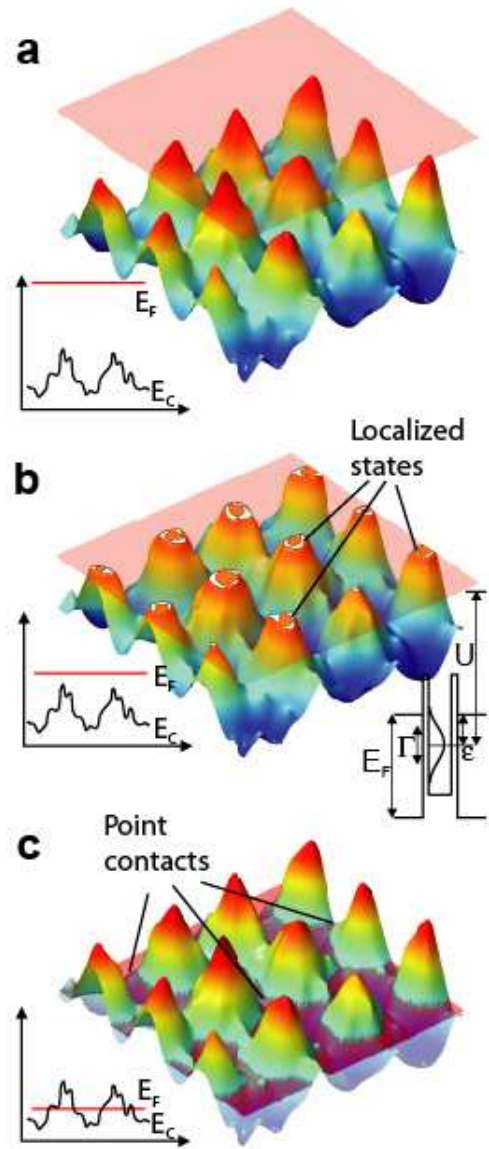}
\caption{Schematic of background disorder and local moment
formation. While the rapid fluctuations arise from dopant density
fluctuations, the slow quasi-regular fluctuations indicates effect
of strong correlation in the donor layer at large filling. {\bf a,}
large, {\bf b,} intermediate, and {\bf c,} low, carrier density
regimes are shown. Our experimental results conform to the
island-in-sea scenario of Fig.~4b.}
\end{figure*}

\end{document}